\documentclass[aps,pra,showpacs,twocolumn,amsmath,amssymb,superscriptaddress,footinbib]{revtex4}

\usepackage[english]{babel}
\usepackage{latexsym}
\usepackage{graphicx}
\usepackage{subfigure}
\usepackage{epsfig}
\usepackage{amsfonts}
\usepackage{amssymb}
\usepackage{amsmath}
\usepackage{bbm}

\begin{document}

\title{Anomalous molecular dynamics in the vicinity of conical intersections}


\author{Jonas Larson}
\email{jolarson@fysik.su.se}
\affiliation{Department of Physics,
Stockholm University, SE-106 91 Stockholm,
Sweden}
\affiliation{Institut f\"ur Theoretische Physik, Universit\"at zu K\"oln, K\"oln, 50937, Germany}
\author{Elham Nour Ghassemi}
\affiliation{Department of Physics,
Stockholm University, SE-106 91 Stockholm,
Sweden}
\affiliation{Institut f\"ur Festk\"orpertheorie, Westef\"alische Wilhelms-Universit\"at, M\"unster, 48149, Germany}
\author{\AA sa Larson}
\affiliation{Department of Physics,
Stockholm University, SE-106 91 Stockholm,
Sweden}

\date{\today}

\begin{abstract}
Conical intersections between molecular electronic potential surfaces greatly affect various properties of the molecule. Molecular gauge theory is capable of explaining many of these often unexpected phenomena deriving from the physics of the conical intersection. Here we give an example of anomalous dynamics in the paradigm of the $E\times\varepsilon$ Jahn-Teller model, which does not allow a simple explenation in terms of standard molecular gauge theory. By introducing a dual gauge theory, we unwind this surprising behavior by identifying it with an intrinsic spin Hall effect. Thus, this work link knowledge of condensed matter theories with molecular vibrations. Furthermore, via {\it ab initio} calculations the findings are as well demonstrated to appear in realistic systems such as the Li$_3$ molecule.  
\end{abstract}

\maketitle

{\it Introduction.} -- The most well known example of a gauge theory is the one of a charge particle moving in an electromagnetic field~\cite{jackson}. Denoting the particle mass and charge by $m$ and $q$ respectively, the minimal coupling Hamiltonian reads
\begin{equation}\label{minimal}
\hat{H}=\frac{\left(\hat{\mathbf{p}}-q\mathbf{A}(\hat{\mathbf{x}})\right)^2}{2m}+q\Phi(\hat{\mathbf{x}}),
\end{equation}
with $\mathbf{A}(\hat{\mathbf{x}})=\left(A_x(\hat{\mathbf{x}}),A_y(\hat{\mathbf{x}}),A_z(\hat{\mathbf{x}})\right)$ the vector potential and $\Phi(\hat{\mathbf{x}})$ the scalar potential. The curl of the vector potential and the gradient of the scalar potential give respectively the gauge invariant magnetic and electric fields. In the seminal work~\cite{mead}, it was demonstrated how the Hamiltonian describing vibrations within molecules can be put in a form (\ref{minimal}) by expressing it in an adiabatic basis. Similar ideas have since then been put forward in other fields~\cite{jonas1}. In contrast to the gauge theory of electromagnetism, such a synthetic gauge theory can become non-Abelian in which the vector potential components $A_\alpha(\hat{\mathbf{x}})$ ($\alpha=x,\,y,\,z$) are mutually non-commuting matrices~\cite{molgauge}. Accompanying non-Abelian properties give rise to novel phenomena well studied in high energy or condensed matter theories. Certain cases arise in solid state materials with spin-orbit couplings, resulting in among others intrinsic spin or anomalous Hall effects~\cite{she} or Majorana fermion quasi particles.  

In molecular and chemical physics, the gauge formalism is particularly effective for understanding molecular dynamics in the vicinity of {\it conical intersections} (CI). Dating back to the 50'th it has been known that the electronic wave function changes sign upon encircling a CI~\cite{LH}, which in terms of the gauge theory can be understood as geometric phase effect. The resulting synthetic magnetic field is everywhere zero except at the CI where it diverges, and the geometric phase is proportional to the magnetic flux through the CI. The simplest model describing this effect is the $E\times\varepsilon$ Jahn-Teller (JT) Hamiltonian $\hat{H}_{E\times\varepsilon}$~\cite{molgauge,jt}. In this work we demonstrate that standard molecular gauge theories, when accounted for carelessly, do not predict certain behavior even in the most simple case of the $E\times\varepsilon$ Jahn-Teller model. Apart at the CI, the effective Lorentz forse is everywhere zero, and furthermore the electric and the classical forces (stemming from the adiabatic potential surfaces) are along the radial direction. Despite this, simulations show how a localized wave packet may be accelerated in the angular direction. This anomalous behaviour is identified as an intrinsic spin Hall effect and is readily resolved in a dual representation of the problem. The dual counterparts of the vector and scalar potentials are manifestly non-Abelian, and the accelerated angular motion can be understood as deriving from a non-vanishing dual Lorentz force. While the phenomenon is easily understood for the linear $E\times\varepsilon$ JT model, more realistic settings do not allow for an analytic dual analysis. From {\it ab initio} calculations of the energy potentials of the Li$_3$ molecule we perform wave packet simulations and verify the effect also for experimentally relevant systems.

{\it Molecular gauge theory of the linear $E\times\varepsilon$ Jahn-Teller model.} -- For the linear $E\times\varepsilon$ JT model, the doubly degenerate vibrational mode with frequency $\omega$ will couple the two degenerate diabatic electronic states according to
\begin{equation}\label{Exe}
\hat{H}_{E\times\varepsilon}=\omega\!\left(\frac{\hat{P}_x^2}{2}+\frac{\hat{P}_y^2}{2}+\frac{\hat{Q}_x^2}{2}+\frac{\hat{Q}_y^2}{2}\right)+k\!\left(\hat{Q}_x\hat{\sigma}_x+\hat{Q}_y\hat{\sigma}_y\right)\!,
\end{equation}
where $\hat{P}_{x,y}$ and $\hat{Q}_{x,y}$ are two sets of conjugate variables with $[\hat{Q}_{x,y},\hat{P}_{x,y}]=i$, and $k$ is the JT coupling strength. The $\sigma$-operators are the standard Pauli matrices; $\hat{\sigma}_x=|1\rangle\langle 2|+|2\rangle\langle 1|$, $\hat{\sigma}_y=i\left(|1\rangle\langle 2|-|2\rangle\langle 1|\right)$, $\hat{\sigma}_z=|2\rangle\langle 2|-|1\rangle\langle 1|$, with $|1\rangle$ and $|2\rangle$ denoting the internal diabatic electronic states. $\hat{H}_{E\times\varepsilon}$ possesses a continuous $U(1)$ symmetry, conservation of $z$ angular momentum $\hat{J}_z=\hat{L}_z+\frac{1}{2}\hat{\sigma}_z$ where $\hat{L}_z$ is the regular angular momentum operator, and a discrete $Z_2$ symmetry characterized by the operator $\hat{T}$ defined as $\hat{T}\hat{\mathbf{x}}\hat{T}^{-1}=-\hat{\mathbf{x}}$ and $\hat{T}\hat{\sigma}_\alpha\hat{T}^{-1}=-\hat{\sigma}_\alpha$. The second of these is the dual version of the regular time-reversal operator and $\hat{T}$-invariance implies a Kramer's' degeneracy. As a final remark, throughout this work we will use atomic units.

In the standard gauge theory of molecules~\cite{molgauge}, the $x$- and $y$- components of the synthetic gauge potential are $\hat{A}_{\alpha}^{(kl)}=-i\langle\psi_k|\partial_{Q_\alpha}|\psi_l\rangle$, 
where the matrix-indices $k$ and $l$ are either $-$ or $+$, and the adiabatic states $|\psi_+\rangle=\left[\exp(-i\varphi/2)|1\rangle-\exp(i\varphi/2)|2\rangle\right]/\sqrt{2}$ and  $|\psi_-\rangle=\left[\exp(-i\varphi/2)|1\rangle+\exp(i\varphi/2)|2\rangle\right]/\sqrt{2}$, with the $Q$-dependent phase $\varphi=\arctan\left(Q_y/Q_x\right)$. The adiabatic states diagonalize the JT coupling part in Eq.~(\ref{Exe}) with corresponding adiabatic energy potentials $\hat{V}_\pm(\hat{Q}_x,\hat{Q}_y)=\omega\left(\hat{Q}_x^2+\hat{Q}_y^2\right)/2\pm k\sqrt{\hat{Q}_x^2+\hat{Q}_y^2}$. The overall phase ambiguity of the adiabatic states $|\psi_\pm\rangle$ represents the gauge freedom. The gauge independent magnetic field can be expressed in terms of the field tensor as $\hat{\mathbf{B}}_i=\frac{1}{2}\varepsilon_{ijk}\hat{\mathbf{F}}_{jk}$, where $\varepsilon_{ijk}$ is the Levi-Cevita symbol and
\begin{equation} \hat{\mathbf{F}}_{jk}=\partial_j\hat{A}_k-\partial_k\hat{A}_j-i[\hat{A}_j,\hat{A}_k].
\end{equation}
The last part of the field tensor derives from the matrix structure of the gauge potential, and whenever the two components $\hat{A}_x$ and $\hat{A}_y$ do not commute the system is said to be non-Abelian. In terms of the $E\times\varepsilon$ JT model, 
\begin{equation}
\begin{array}{c}
\hat{A}_x=\displaystyle{\frac{\hat{Q}_y}{\hat{Q}_x^2+\hat{Q}_y^2}\hat{\sigma}_y},\hspace{1cm}\hat{A}_y=\displaystyle{-\frac{\hat{Q}_x}{\hat{Q}_x^2+\hat{Q}_y^2}\hat{\sigma}_y},\\ \\
\hat{\Phi}=\displaystyle{\frac{1}{8\left(\hat{Q}_x^2+\hat{Q}_y^2\right)}},
\end{array}
\end{equation}
resulting in an Abelian synthetic magnetic field vanishing everywhere except at the CI at $(Q_x,Q_y)=(0,0)$, where it becomes singular; ergo also the Lorentz force must strictly disappear in the punctured $Q_xQ_y$-plane $\mathbb{R}\backslash(0,0)$. Moreover, since the adiabatic states $|\psi_\pm\rangle$ are independent on $\omega$ and $k$, it follows that also the magnetic field is independent on the system parameters, which reflects the geometric phase $\gamma=\pi$ irrespective of system parameters. The electric and classical forces ($\propto-\nabla\hat{\Phi}$ and $\propto-\nabla \hat{V}$ respectively) are pointing both in the radial direction, i.e. $[\hat{L}_z,\hat{V}]=[\hat{L}_z,\hat{\Phi}]=0$. 

\begin{figure}[h]
\centerline{\includegraphics[width=8cm]{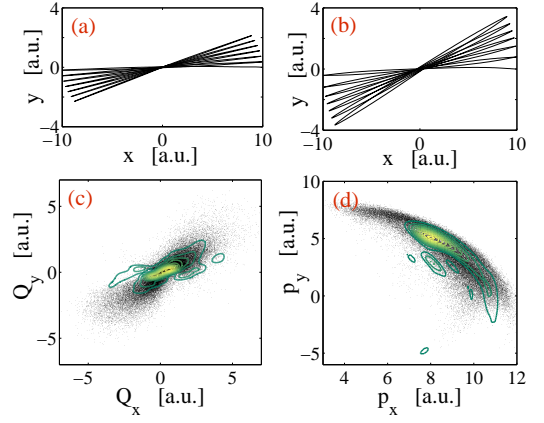}}
\caption{(Color online) In (a) the time evolution of the quantum expectation values ($x(t),y(t)$) are displayed, while (b) shows the corresponding semiclassical averaged trajectories obtained using the TWA approach. The final quantum density distributions in real and momentum space are depicted with contour-lines in (c) and (d), while the dots show the semi-classical distributions.} \label{fig1}
\end{figure}
 
{\it Anomalous dynamics in the $E\times\varepsilon$ Jahn-Teller model.} -- If we release a localized wave packet with initial expectation values $(p_{x}(0),p_{y}(0))=(0,0)$ and $(x(0),y(0))=(x_0,0)$, it appears reasonable to expect $y(t)=0$ since the effective Lorentz force is strictly zero outside the CI and should not affect the dynamics as long as the wave packet is far from the CI. As already pointed out, such behavior is not what is found. We numerically demonstrate the anomalous wave packet propagating, using the split-operator method~\cite{split} with an initial Gaussian state,
\begin{equation}
\psi(Q_x,Q_y,0)=\!\left(\frac{1}{\pi\sigma^2}\right)^{1/4}\!\exp\left[-\frac{(Q_x-x_0)^2+Q_y^2}{2\sigma^2}\right]\!|1\rangle,
\end{equation}
with $x_0=10$ a.u. and the width $\sigma=1$ a.u. corresponding to a coherent state given the frequency of the harmonic oscillators. The wave packet is propagated until $t=15\,000$ a.u., and the parameter values, $k=0.01$ a.u. and $\omega=0.02$ a.u. The resulting trajectories of the expectation values $(x(t),y(t))$ are shown in Fig.~\ref{fig1} (a). Directly after the propagation has started, a positive $y(t)$ component builds up. As soon as the wave packet has crossed into negative $Q_x$, the effective force is reversed and $y(t)$ decreases again and at the first turning point, $x(t)\approx-10$ a.u., it has become negative $y(t)<0$. For continued evolution, a ``bending'' of the harmonic motion takes place in which vibrational excitations in the $x$-mode is swapped into excitation in the initially empty $y$-mode. This can be seen as a non-zero transverse current in the $\phi$-direction. The density $|\psi(Q_x,Q_y,t)|^2$ and its corresponding momentum density $|\phi(P_x,P_y,t)|^2$ of the final wave packet are displayed with solid contour lines in Fig.~\ref{fig1} (c) and (d). During the propagation the wave packet stays localized both in real and in momentum space, while for even longer times the wave packet, due to anharmonicities originating from the JT coupling, will spread out and cover more or less the entire accessible phase space~\cite{com1}. To exclude some tail-effects, i.e. there is always a small portion of the wave packet being non-zero at the CI, we have varied the width of the initial state and verified that the effect at short times is indeed independent of the initial wave packet width.

Before presenting an explanation of the unexpected behavior in terms of a Hall effect, we first make the phenomenon more transparent by deriving a set of semi-classical equations of motion, similar to those of Wong describing a classical color-charged particle in a non-Abelian field~\cite{wong},
\begin{equation}\label{wongeq}
\begin{array}{l}
\dot{x}=\omega p_x,\\
\dot{p}_x=-\omega x-ks_x,\\
\dot{y}=\omega p_y,\\
\dot{p}_y=-\omega y-ks_y,\\
\dot{s}_x=kys_z,\\
\dot{s}_y=-kxs_z,\\
\dot{s}_z=k\left(xs_y-ys_x\right),
\end{array}
\end{equation}  
where we recognize $s_\alpha=\langle\hat{\sigma}_\alpha\rangle$ with $\alpha=x,\,y,\,z$, and the dot indicates time derivative. Note that we denote the above semi-classical variables the same as the quantum expectation values, even though they are not equivalent in a strict sense. Indeed, the above set of equations does not take higher order quantum correlations into account, i.e. the equations are truncated and do not consider products like $\dot{(xs_z)}$, $\dot{(ys_x)}$ and so on. From Eq.~(\ref{wongeq}) we have $\ddot{y}=-\omega^2y-\omega ks_y$, and after linearizing for short times we find $y\sim k^2t^3$. Clearly, a build-up of a non-zero $y$ must derive from coupling of $p_y$ to the spin degree-of-freedom. 

The semi-classical approach is compared with the full quantum wave packet propagation using the truncated Wigner approximation (TWA), which has turned out to often reproduce the quantum dynamics surprisingly well~\cite{twa}. Thus, we randomly sample initial values of positions and momentum according to the probability distributions $|\psi(Q_x,Q_y,0)|^2$ and $|\phi(P_x,P_y,0)|^2$ and propagate them using Eqs.~(\ref{wongeq}). The expectation values displayed in Fig.~\ref{fig1} (b) are approximated with the averages over all trajectories. The final semi-classical distributions obtained from $N=50\, 000$ sampled trajectories are depicted in (c) and (d). It is found that this method somewhat overestimates the transverse motion, but still captures the quantitative features. This discrepancy is most likely due to large quantum fluctuations in the spin variables which is not captured in the above semi-classical analysis. As initial values for the spin isovector we used $S=(s_x,s_y,s_z)=(0,0,1)$ for all sampled trajectories, i.e. we have not taken initial quantum fluctuations into account for the internal two-level system~\cite{com2}. 

In addition, for both the quantum and semi-classical approaches, we have numerically verified that the position in the transverse direction, i.e. the build-up of a transverse current, scales as $\sim k^2t^3$ for short times as predicted by the linearized semi-classical equations. For longer times, due to the interplay between external and internal dynamics, the evolution becomes rather complex, which is also known already for free particles subjected to the same type of coupling for which the problem is analytically tractable~\cite{jonas3}. 

It is important to note that for the parameters $k$ and $\omega$, the ratio $\omega/k$ is larger than that obtained from {\it ab initio} calculations of the Li$_3$ molecule. 
As illustrated below for the dynamics on the two lower states of Li$_3$, the anharmonicity induced by the CI greatly destroys the harmonic motion and therefore also the regular structure of Fig.~\ref{fig1}.
 
{\it Generalized gauge theory of the linear $E\times\varepsilon$ Jahn-Teller model.} -- 
With simple arguments of the standard molecular gauge theory we were not able to explain the anomalous dynamics captured in the numerics. From the semi-classical analysis it followed that the origin of the phenomenon must be found in some sort of spin-orbit coupling. This will become more clear by introducing a dual representation of the problem, where one sees how the anomalous evolution can be identified as a version of the intrinsic spin Hall effect. To this end, let us rewrite the JT Hamiltonian as
\begin{equation}\label{Exe2}
\hat{H}_{E\times\varepsilon}\!=\!\omega\!\left[\!\frac{\left(\hat{Q}_x\!-\!\tilde{A}_x\right)^2}{2}\!+\!\frac{\left(\hat{Q}_y\!-\!\tilde{A}_y\right)^2}{2}+\frac{\hat{P}_x^2}{2}+\frac{\hat{P}_y^2}{2}\!\right]\!+\hat{\Phi}\!,
\end{equation}
where $\tilde{A}_{x,y}=-\frac{k}{\omega}\hat{\sigma}_{x,y}$ are the two components of the dual gauge vector potential, and $\hat{\Phi}=k^2/\omega$ the dual scalar potential. In this representation, the gauge potentials are constant but, due to their non-Abelian character, give rise to a non-trivial dual synthetic magnetic field and a dual Lorentz force,
\begin{equation}\label{bfield}
\begin{array}{lll}
\displaystyle{\tilde{\mathbf{B}}_z=-i[\tilde{A}_x,\tilde{A}_y]\!=\!2\frac{k^2}{\omega^2}\hat{\sigma}_z}, & & \displaystyle{\mathbf{F}_{dLor}\!=\!2\frac{k^2}{\omega^2}\hat{\sigma}_z(-\hat{Q}_y,\hat{Q}_x)}.
\end{array}
\end{equation}
The interpretation of Eq.~(\ref{Exe2}) is a particle with two internal states moving in a harmonic potential influenced by a constant state-dependent magnetic field; if the particle is in the $|1\rangle$-state it sees a field pointing in the negative $z$-direction and the opposite holds for state $|2\rangle$. Unlike the synthetic magnetic field in the position representation, the dual magnetic field does depend on both $\omega$ and $k$. Consequently, it can not be seen as some sort of Fourier transform of the latter. Furthermore, put in a momentum representation, the JT coupling is nothing but a Rashba spin-orbit coupling~\cite{rashba} frequently appearing in condensed matter theories, such as the intrinsic anomalous and spin Hall effects~\cite{she}. It is important to remember that the spatial angular momentum $\hat{L}_z$ is not conserved alone, only the total angular momentum $\hat{J}_z$ is conserved, and it is this fact which makes transverse angular momentum possible. The Lorentz force $\mathbf{F}_{dLor}$ acts in the dual space causing a deviation of the momentum. The build-up of a non-zero $p_y$ propagates to the position $y$. This section clarifies that in the alternative picture, the syntethic magnetic field indeed predicts the anomalous dynamics as a result of the Rashba-form of the coupling.

\begin{figure}[h]
\centerline{\includegraphics[width=8cm]{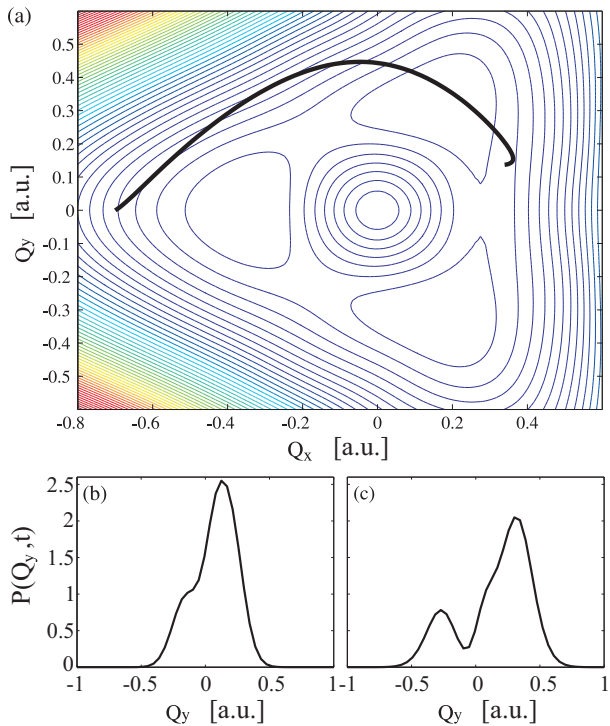}}
\caption{(Color online) The upper plot (a) shows the lower adiabatic energy potential surface of the Li$_3$-trimer (contour lines) together with the quantum expectation values $x(t)$ and $y(t)$ (thick black line). The lower figures display the distribution $P(Q_y,t)$ at time instants, $t\approx410$ a.u. (b) and $t\approx820$ a.u. (c). The initial state is a Gaussian with $y(0)=p_x(0)=p_y(0)=0$ a.u. and $x(0)=-0.7$ a.u., and width $\sigma=1/2$ a.u. The final propagation time is approximately 4100 a.u. It is particularly clear how the system, as an appearance of the spin Hall effect, directly builds up a non-zero displacement in the $Q_y$-direction.} \label{fig2}
\end{figure}

{\it Effect in Li$_3$-trimer.} -- In the idealized case of the linearized $E\times\varepsilon$ JT model, the presence of a transverse force is clear from the dual representation. Once higher order terms are taken into account, i.e. the coupling and the potential are no longer linear and quadratic respectively, introducing a dual gauge potential is non-trivial. Nevertheless, we may expect that the dynamics should still be greatly influenced by effects deriving from linear order terms. Thus, as for the above example, the augmentation of a transverse current is taken as a smoking gun for a spin Hall effect which in return can be ascribed some synthetic Lorentz force. 

As a demonstrating example we consider the Li$_3$ molecule which possesses a CI between its two lowest electronic potentials~\cite{li3}. More precisely, the Li$_3$ molecule possesses three vibrational normal modes corresponding to bending, asymmetric stretching, and symmetric stretching modes, which are all taken into account in our analysis. The adiabatic potential energy surfaces are computed with the MOLPRO package~\cite{molpro} using the Multi-Reference Configuration Interaction (MRCI) method with molecular orbitals obtained from the state-averaged Multi-Configuration Self Consistent Field (MCSCF) calculations with an active space consisting of three electrons distributed among fifteen orbitals (all molecular orbitals composed of $2s$ and $2p$ atomic orbitals). The MRCI calculations are performed using the same reference space as for the MCSCF calculations. Excitations out of the core orbitals as well as single and double external excitations are all included. The calculations are carried out using the aug-cc-PVTC basis set~\cite{basis}. By least square fitting of the computed two lowest adiabatic potential energy surfaces to the eigenvalues of the potential part of the JT Hamiltonian, the corresponding JT parameters can be obtained~\cite{diabat}. Here, in addition to the linear terms, the fitted Hamiltonian also contains quadratic and cubic anharmonic terms. The obtained JT parameters are used to determine the adiabatic to diabatic transformation matrix. Then the \textit{ab initio} computed adiabatic potential energy surfaces are transformed into the diabatic potentials and couplings used for the wave packet propagation. A reliable description of the system Hamiltonian is then obtained also far from the conical intersection, where the JT Hamiltonian is no longer a good approximation~\cite{diabat}. The details of the numerical procedures for the Li$3$-calculations, together with corresponding results of the potentials, will be published elsewhere~\cite{elham}. 

The full three-dimensional wave packet propagation is carried out using the Multi-Configurational Time-Dependent Hartree method~\cite{mctdh}. The initial state is a real valued Gaussian with width $\sigma=1/2$ a.u. located at the lower adiabatic potential surface at $(Q_x,Q_y,Q_s)=(-0.7,0,3.2)$ a.u. For this choice, the initial state is, to a good approximation, not in contact with the synthetic flux through the CI. The value $Q_s=3.2$ corresponds to a global minima in this coordinate, and the dynamics should predominantly appear in the $Q_x$ and $Q_y$ coordinates corresponding to bending and asymmetric stretching. The lower potential surface is symmetric with respect to $Q_y=0$ (see Fig.~\ref{fig2}), and, as for the linear $E\times\varepsilon$ JT model, this implies that a non-zero $y(t)$ signals a spin Hall effect. The numerical results are presented in Fig.~\ref{fig2}, showing the trajectory $(x(t),y(t))=(\langle\hat{Q}_x\rangle,\langle\hat{Q}_y\rangle)$ with a thick line in (a), and two snap-shots of the projected distribution $P(Q_y,t)=\int\,dQ_x|\psi(Q_x,Q_y,t)|^2$ in (b) and (c). The instantaneous build-up of a transverse force/current is clearly demonstrated in the figure, and note that the sign of this force is in accordance with our previously introduced dual Lorenz force. This is also demonstrated in the asymmetry of the projected ´distributions already seen at very short times. Due to large anharmonicities, the wave packet rapidly spreads and the effect is therefore only visible for short times. It is interesting to observe that the Li$_3$ molecule is chiral and it is therefor not clear how the present behavior can be credited some broken chirality.

{\it Conclusion.} -- Using knowledge of the dual model of the $E\times\epsilon$ JT Hamiltonian, we demonstrated how one can explain anomalous dynamics in the vicinity of the CI. Such effects are not easily captured in standard molecular gauge theory where especially the synthetic Lorentz force vanishes. In the dual picture, the inherent gauge potentials are non-Abelian and constitute an everywhere non-zero Lorentz force which causes a wave packet far from the CI to deflect from its expected classical trajectory. In condensed matter theories, the same mechanism is essential for; the spin Hall effect, topological insulators, high-T superconductors, Dirac and Majorana fermions, just to mention a few examples. In this respect, the present work bridges two of the main fields of physics, molecular and condensed matter, and our results suggest that known effects in one field can be translated into the other.

Our dual picture analysis is exact in the linear case, but is not apparent when non-harmonicity and an extra vibrational degree-of-freedom are taken into account. Using a full quantum calculation, taking all degrees-of-freedom into account as well as all higher orders in the JT model, we also presented evidences how the intrinsic asymmetry of the Li$_3$ molecule dynamics can be assigned a spin Hall effect.     

\begin{acknowledgements}
We thank Hans Hansson and Erik Sj\"oqvist for valuable discussions. JL acknowledges support from the Swedish research council (VR), Deutscher Akademischer Austausch Dienst (DAAD), and Kungl. Vetenskapsakademien (KVA). {\AA}L acknowledges support from the Swedish research council (VR).
\end{acknowledgements}


\begin{thebibliography}{999}
\bibitem{com1} By ``accessible'' we mean the surface of phase space bounded to the energies of the initial state.
\bibitem{com2} One could include spin fluctuations in the analysis by representing it in terms of spin coherent states, but since we are only interested in the qualitative behavior we do not find this necessary, and furthermore.
\end{thebibliography}
\end{document}